\documentclass[journal]{IEEEtran}
\usepackage{amsmath,amsfonts}
\usepackage{algorithmic}
\usepackage{epsfig, algorithm}
\usepackage{array}
\usepackage{amssymb}
\usepackage{acronym} 
\usepackage[caption=false,font=normalsize,labelfont=sf,textfont=sf]{subfig}
\usepackage{textcomp}
\usepackage{stfloats}
\usepackage{tabularx}
\usepackage{xcolor}
\usepackage{url}
\usepackage{verbatim}
\usepackage{multirow}
\usepackage{acronym}
\usepackage{graphicx}	
\usepackage{stfloats}
\usepackage{soul}
\usepackage{cancel}
\def\BibTeX{{\rm B\kern-.05em{\sc i\kern-.025em b}\kern-.08em
T\kern-.1667em\lower.7ex\hbox{E}\kern-.125emX}}
\usepackage{balance}
\usepackage{multirow}
\usepackage{hhline}
\input{AcronymsListFinal}

\usepackage{fancyhdr}

\begin{document}
\title{Polarization-Aware Antenna Selection for Joint Radar and Communication in XL-MIMO Systems}
\author{Ahmed~Naeem, Liza Afeef, Hüseyin~Arslan~\IEEEmembership{Fellow,~IEEE}
\thanks{The authors are with the Department of Electrical and Electronics
Engineering, Istanbul Medipol University, Istanbul, 34810, Turkey (email: ahmed.naeem@std.medipol.edu.tr, lizaafeef92@yahoo.com, huseyinarslan@medipol.edu.tr).}
\\This work has been submitted to the IEEE for possible publication. Copyright may be transferred without notice, after which this version may no longer be accessible.}
\maketitle
\begin{abstract}
A key challenge in dual-polarized multiplexing for joint radar and communication (JRC) systems is cross-polarization (cross-pol) leakage caused by depolarization. In conventional MIMO systems, depolarization arises solely from the channel; however, in XL-MIMO systems, non-stationary properties of the array cause additional polarization shifts at each antenna element, further degrading JRC performance. This paper introduces a channel model incorporating polarization shifts due to the propagation channel and antenna elements in the near-field. We also propose an antenna selection (AS) scheme that dynamically chooses antennas based on polarization imbalance and cross-pol leakage, enhancing spectral efficiency, symbol error rate, and radar detection probability. Simulations show that the proposed AS significantly outperforms traditional methods, providing scalable benefits for XL-MIMO JRC systems.
\end{abstract}
\begin{IEEEkeywords}
Channel model, dual-polarization, joint radar and communication, multiplexing, polarization, XL-MIMO.
\end{IEEEkeywords}
\section{Introduction}
\par \Ac{XL}-\ac{MIMO} is a fundamental enabler for next-generation wireless networks, particularly in \ac{JRC} and radar applications. Unlike traditional \ac{FF} \ac{MIMO}, \ac{NF} \ac{XL}-\ac{MIMO} utilizes large-scale, densely packed antenna arrays, leading to spherical wavefront propagation, high resolution sensing, and enhanced spatial multiplexing. Due to the \ac{NF} property, this transition from planar to spherical wavefront significantly impacts channel modeling and system design \cite{wang2023extremely, cui2022near}. \ac{XL}-\ac{MIMO} deployment also introduces key electromagnetic characteristics, such as spatial non-stationarity \cite{de2020non}, mutual coupling \cite{svantesson2001mutual}, and polarization effects \cite{yuan2021electromagnetic}, which~are crucial for advanced radar, autonomous systems, and future wireless networks.
\par \Ac{NF} \ac{XL}-\ac{MIMO} introduces spherical rather than planar wavefront propagation, resulting in distance-dependent signal variations that impact both sensing and communication accuracy \cite{wang2023extremely,cui2022near}. Additionally, the non-stationary channel behavior across the array causes polarization shifts and varied \acp{AoA}, complicating traditional beamforming techniques \cite{de2020non}. While these challenges pose difficulties, they also offer new opportunities to exploit spatial and polarization diversity, improving sensing accuracy and interference management. However, the shared spectrum in \ac{JRC} applications increases the complexity of interference mitigation, requiring advanced signal processing techniques.
\par Polarization diversity and multiplexing play a key role in multi-user separation, interference suppression, and target detection for \ac{JRC}. Modern \ac{DP} antenna arrays enable the simultaneous transmission and reception of signals on orthogonal polarization states, improving \ac{SE} and spatial multiplexing \cite{10458884}. However, polarization-dependent channel distortions in \ac{NF} XL-\ac{MIMO}, such as depolarization and polarization misalignment, lead to \ac{XPD}, which degrades communication reliability and sensing accuracy, necessitating advanced polarization-aware interference mitigation techniques \cite{yuan2021electromagnetic}.
\par In \ac{DP} XL-\ac{MIMO}, radar and communication signals are transmitted on orthogonal polarization states to enable simultaneous operation within a shared spectrum \cite{10458884}. However, channel depolarization, multipath effects, and polarization misalignment cause polarization leakage, leading to \ac{XPD} and degrading system performance. This results in higher \ac{BER} for communication and lower detection probability, \ac{$P_d$}, in radar sensing, requiring polarization-aware beamforming, adaptive filtering, and \ac{AS} to mitigate interference and maintain robust \ac{JRC} performance \cite{yuan2021electromagnetic}.
\subsection{Prior Works}
\Ac{JRC} systems can be realized through three major architectures: (a) spectral overlap, (b) cognitive coexistence, and (c) functional coexistence \cite{zheng2019radar}. This work focuses on spectral overlap, where \ac{JRC} operates within the same frequency spectrum, making interference mitigation a critical challenge. Ensuring satisfactory performance for both sensing and communication functions requires advanced interference suppression techniques that balance complexity, \ac{SE}, and system adaptability.
\par Several interference mitigation techniques have been proposed for \ac{JRC} coexistence. The work in \cite{zheng2017adaptive} introduces two sparsity-based interference mitigation algorithms using compressed sensing and \ac{AN} minimization for uncoordinated \ac{JRC} coexistence. However, AN-based methods suffer from high computational complexity, making them impractical for large-scale systems. Similarly, \cite{li2019interference} presents an alternating minimization scheme for radar-induced interference suppression, but it faces non-convexity issues and relies on sparsity assumptions, which may not hold in high-noise or strong multipath environments.
\par Signal processing techniques have also been explored for spectrum sharing between \ac{MIMO} radar and wireless communication. The work in \cite{deng2013interference} proposes coherent \ac{MIMO} radar and cognitive radio-based interference mitigation, but it requires accurate channel estimation and suffers from high computational complexity due to advanced spatial filtering. Similarly, in \cite{ciuonzo2016intrapulse}, the radar system periodically senses the spectrum, allowing the cellular \ac{BS} to transmit only in sidelobe regions while maintaining a minimum \ac{INR}. However, this approach restricts spectrum utilization and imposes strict power constraints on communication systems. Precoder-based designs leverage interference \ac{CSI} to suppress interference, particularly when radar has primary spectrum access \cite{liu2019interfering}. \Ac{SVD}-based null-space precoding achieves zero-forced interference cancellation \cite{mahal2017spectral}, further optimized in cluttered environments \cite{li2017joint, liu2018mimo}. However, these methods depend on precise interference CSI estimation, requiring additional training signals, increasing computational complexity, and spectrum overhead. To address this, \cite{rao2020probability} suggests prioritizing target detection over CSI acquisition, emphasizing sensing performance in spectrum-sharing scenarios.
\par The convergence of \ac{XL}-\ac{MIMO} and JRC has been investigated for various applications. In \cite{10520715}, an iterative beamforming approach is formulated for multi-target detection in \ac{NF} JRC systems. Similarly, \cite{10579914} proposes a \ac{NF} JRC framework where a BS simultaneously serves multiple communication \acp{UE} while performing target sensing. Additionally, \cite{10772413} presents three beamforming schemes aimed at maximizing the \ac{SINR}. While these works explore beamforming and sensing strategies, they assume \ac{LoS} channels between the \ac{XL}-\ac{MIMO} array and targets, which may not always exist due to environmental scatterers and obstructions. To address this, \cite{9737357} investigates networked sensing through information fusion, providing a robust sensing architecture for \ac{NLoS} environments.
\par Despite extensive research on JRC coexistence, precoding-based interference mitigation, and \ac{XL}-\ac{MIMO}-enabled \ac{JRC}, none of the existing works have explored \ac{NF} \ac{XL}-\ac{MIMO} polarization effects for interference suppression. In particular, \ac{DP} interference in \ac{NF} \ac{XL}-\ac{MIMO} remains unaddressed, as conventional approaches assume ideal polarization orthogonality and overlook non-stationary polarization behavior across the array. Furthermore, existing \ac{AS} strategies in \ac{JRC} systems primarily focus on spatial beamforming
\cite{10437147}, neglecting polarization-aware selection mechanisms that can significantly enhance signal separation and interference rejection.
\subsection{Contributions}
To bridge these gaps, this paper proposes a polarization-based interference management framework for NF \ac{XL}-\ac{MIMO} \ac{JRC} networks, leveraging adaptive \ac{AS} to mitigate \ac{XPD} and optimize sensing-communication coexistence. The key contributions of this work are as follows:
\begin{itemize}
    \item We propose a polarization-aware interference mitigation technique tailored for \ac{NF} \ac{XL}-\ac{MIMO} \ac{JRC} networks. Unlike traditional methods that rely solely on spatial filtering or time-frequency multiplexing, our approach exploits polarization diversity to suppress \ac{XPD}, improving the coexistence of sensing and communication functions.
    \item A detailed polarization model is developed to capture polarization variations across the array, considering distance-dependent depolarization, non-stationary behavior, and \ac{NF} wavefront effects. This model enables more accurate precoding, beamforming, and interference suppression strategies in practical XL-\ac{MIMO} \ac{JRC} systems.
    \item We propose a dynamic \ac{AS} mechanism that optimally selects antennas based on polarization characteristics and interference power. This selection strategy enhances signal separation, improves beamforming efficiency, and minimizes cross-polarization (cross-pol) leakage, making it particularly suitable for NF JRC deployments.
    \item The proposed approach is evaluated using key performance indicators, including \ac{SER} for communication, \ac{$P_d$} for radar sensing, and computational complexity. Our results demonstrate the trade-offs between interference mitigation, signal quality, and system overhead, highlighting the feasibility of polarization-based interference suppression in \ac{XL}-\ac{MIMO}.
    \item Numerical simulations and analytical studies validate the effectiveness of our polarization-aware \ac{JRC} framework, showing significant improvements in interference mitigation, \ac{SE}, and system reliability. These findings confirm the practical applicability of polarization-based methods for next-generation aerospace and electronic systems.
\end{itemize}

\subsection{Organization and Notations}
The rest of the paper is organized as follows. Section II discusses the system model and the proposed \ac{XL}-\ac{MIMO} channel model, Section III introduces the problem formulation, while Section IV explains the proposed \ac{AS}. The system~analysis is discussed in Section V and simulation results are presented~in Section VI. Section VII includes conclusion and future work.
\begin{figure}
\centering 
\resizebox{1\columnwidth}{!}{
\includegraphics{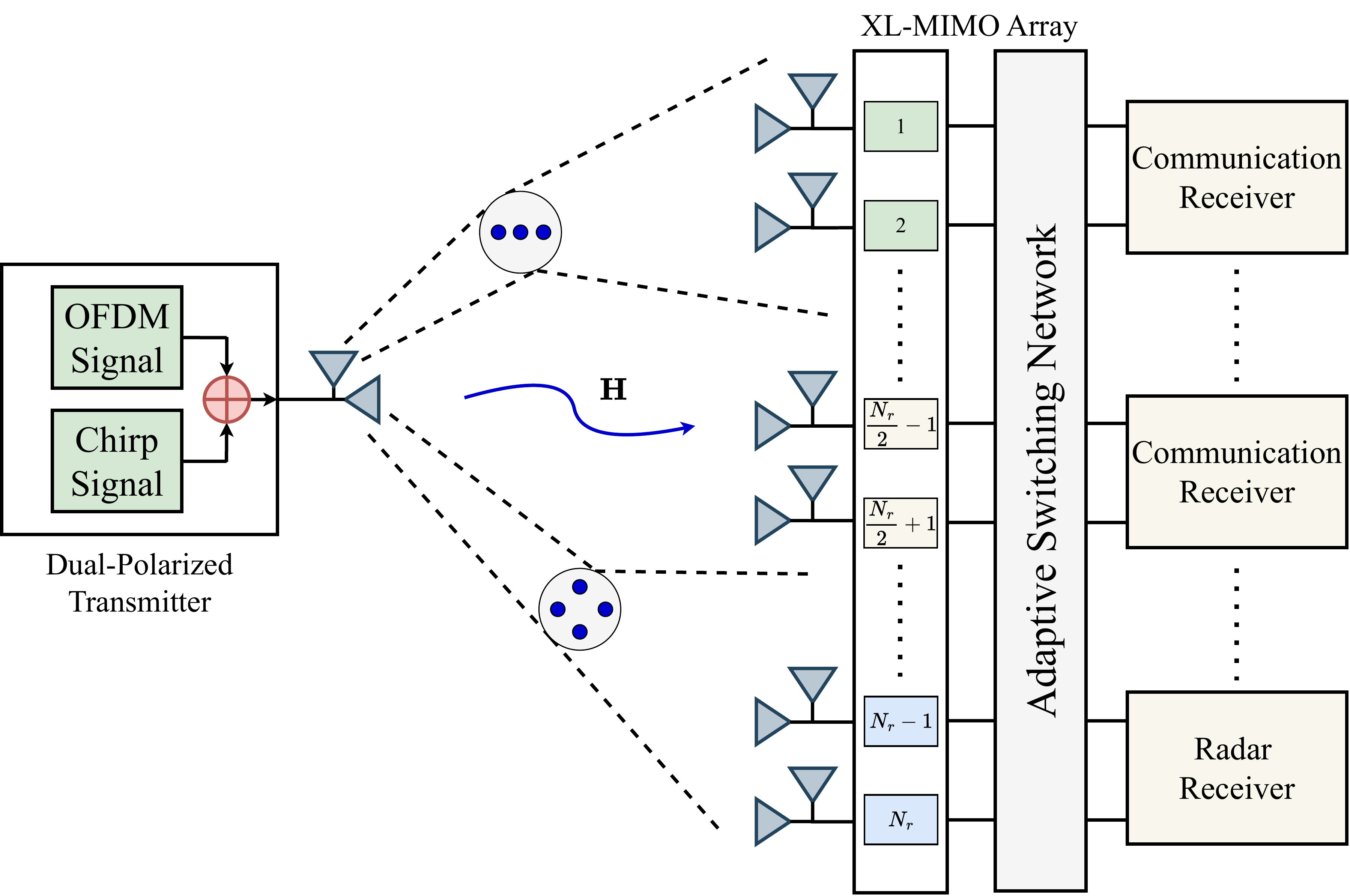}}
\caption{Proposed system model.}
\label{figi1}
\end{figure}

\section{System Model} \label{syss}
\par Assume a \ac{NF} \ac{XL}-\ac{MIMO} \ac{JRC} system as shown in Fig.~\ref{figi1}, where a \ac{DP} \ac{XL} array with $N_r$ antenna elements at a \ac{BS} simultaneously serves single-\ac{DP} antenna communication \acp{UE}~and performs radar sensing. Unlike conventional \ac{FF} \ac{MIMO}, where signals are assumed to have planar wavefront, the large aperture and \ac{NF} propagation in \ac{XL}-\ac{MIMO} lead to spherical wavefront effects. To achieve spectral efficient \ac{JRC} operation, radar and communication signals share the same frequency resources, employing \ac{DP} transmission to enable multiplexing. The system utilizes an adaptive switching network with \ac{AS} mechanism that dynamically activates subset of antennas based on polarization characteristics and interference conditions.
\subsection{Signal and Channel Model}
The transmitted \ac{DP} signal is decomposed into two orthogonal polarized components i.e., horizontal (H) and vertical (V) utilized for simultaneous communication and sensing  to enable efficient multiplexing, which is represented as
\begin{equation}\label{eq7}
    \textbf{E}=\begin{bmatrix}
        E^H\\
        E^V
    \end{bmatrix}=\begin{bmatrix}
    cos\delta \\
    sin\delta ~.~ e^{j\vartheta}
    \end{bmatrix},
\end{equation}
where \ac{SOP} of any \ac{DP} signal is denoted by the amplitude ratio $\delta$ and phase difference $\vartheta$, while $\delta \in [0,\pi/2]$ and $\vartheta \in [0,2 \pi]$, \cite{7794760}. Moreover, $E^H$ is the \ac{OFDM} signal, and $E^V$ is the chirp signal. Due to the propagation channel, the \ac{SOP} gets corrupted by the noise and the depolarization effect causing \ac{XPD}. This leads to the change of the \ac{SOP} and the relationship between the polarization components.
\par The \ac{DP} transmitted signal after it strikes from objects in the environment is received by the common \ac{JRC} receiver. For the scenario mentioned in this paper, the received composite signals at the H \textit{(communication receiver)} and V \textit{(radar receiver)} polarized antennas can be expressed as
\begin{equation} \label{yyy}
\begin{split}
    & \textbf{y}(t)=\mathbf{H}~\textbf{E}(t)+\textbf{w}(t), \\ 
    & \begin{bmatrix}
\textbf{y}^{V}(t)\\ 
\textbf{y}^{H}(t)
\end{bmatrix}=\mathbf{H}\, .\, \begin{bmatrix}
E^H (t)\\ 
E^V (t)
\end{bmatrix}+\begin{bmatrix}
\textbf{w}^H(t) \\
\textbf{w}^V(t)
\end{bmatrix},
\end{split}
\end{equation}
where $\textbf{w}(t)\in \mathbb{C}^{2N_r \times 1}$ is the complex additive white Gaussian noise, $\mathcal{C N}\left(0, \sigma_c^2\right)$, and $\mathbf{H} \in \mathbb{C}^{2N_r \times 2}$ is the polarization-dependent channel matrix. In \ac{XL}-\ac{MIMO}, each antenna element experiences different polarization shifts, leading to spatially varying interference. So, each $n$-$th$ receiver antenna in the \ac{VR} has its own polarization effect as
\begin{equation}\label{depol}
\begin{aligned}
    y^V_n(t) = h^n_{V_r V_t} E^V(t) + h^n_{V_r H_t} E^H(t) + w^V_n(t)\\
y^H_n(t) = h^n_{H_r H_t} E^H(t) + h^n_{H_r V_t} E^V(t) + w^H_n(t)
\end{aligned}
\end{equation}
where, the polarization leakage $h^n_{H_r V_t}, h^n_{V_r H_t}$ is different for each antenna element due to the array’s spatial variation, and the \ac{VR} determines which antennas contribute more to interference.
\par The polarization-dependent channel matrix in NF propagation is given as
\begin{equation}
\mathbf{H}=\left[\begin{array}{ll}
\textbf{h}_{V_rV_t} & \textbf{h}_{V_r H_t} \\
\textbf{h}_{H_rV_t} & \textbf{h}_{H_rH_t}
\end{array}\right]
\end{equation}
where $\textbf{h}_{(V_r/H_r)(H_t/V_t)}$ is the channel gain between the receiver and the transmitter.
In conventional polarized \ac{MIMO} systems, the polarized channel follows the model in \cite{castellanos2023linear}, however, considering the extension to XL \ac{MIMO} with NF effect, the $L$ tap channel model becomes
\begin{equation} \label{main}
\mathbf{H}=\sum_{\ell=1}^{L}\overline{\mathbf{J}}_{\ell, \mathrm{r}}^{\mathsf{T}}\begin{bmatrix}\left(\beta_{\ell}\mathrm{a}_{\mathrm{r}}(\phi_{\ell,\mathrm{r}},d_{\ell,r})\odot u_\ell \right)\otimes\mathbf{Q}_{\ell}(\varphi_{\ell})\mathbf{X}_{\ell}\end{bmatrix}\overline{\mathbf{J}}_{\ell, \mathrm{t}},
\end{equation}
where $\beta_{\ell}$ is the complex path gain for the $\ell$-$th$ path. Moreover, each component of \eqref{main} is explained in detail in next subsection.

\subsection{Array and Polarization Model}
The impact of antenna array configurations and polarization on signal propagation is discussed here, focusing on the co-polarized (co-pol) and cross-pol components in XL-\ac{MIMO}.
\subsubsection{Steering Vector}
The steering vector is given as
\begin{equation}
\begin{aligned}
    \mathrm{a}_{\mathrm{r}}(\phi_{\ell,\mathrm{r}},d_{\ell,r})=\:\frac{1}{\sqrt{N_r}} & [e^{-j\frac{2\pi}{\lambda}(d_{\ell,r}(1)-\bar{d}_{\ell,r})},\cdots, \\
    & e^{-j\frac{2\pi}{\lambda}(d_{\ell,r}(N_r)-\bar{d}_{\ell,r})}]^{H} ,
\end{aligned}
\end{equation}
where $ \phi _{\ell,r}$ represent the \ac{AoA} for the $\ell$-$th$ path at receiver, and $\bar{d}_{\ell,r}$ represent the distance of the $\ell$-$th$ scatterer from the center of the antenna array of receiver. Moreover, $d_{\ell,r}(n)$ is
\begin{equation}
    d_{\ell,r}(n)=\sqrt{\bar{d}_{\ell,r}+\delta_{n}^2d_{\text{array}}^2-2\bar{d}_{\ell,r}\delta_{n}d_{\text{array}}\sin\phi_{\ell,r}}
\end{equation}
represents the distance of the $\ell$-$th$ scatterer to the $n$-$th$ receive antenna element, $d_{\text{array}}$ is inter-element spacing, and $\delta_{n}= \frac {2n- N_r- 1}2$ with $n=1,2,\cdots,N_r$. 
\subsubsection{Antenna Gains} From \eqref{main}, $\overline{\mathbf{J}}_{\ell, \mathrm{r}}$ and $\overline{\mathbf{J}}_{\ell, \mathrm{t}}$ are the receiver and transmitter antenna gains for each $\ell$-$th$ path, respectively. Let $\overline{\mathbf{J}}_\ell \in \mathbb{C}^{2N_r \times 2}$ denote the block diagonal matrix, $\overline{\mathbf{J}}_\ell=\text{blkdiag}\left(\mathbf{J}_1(\phi_{\ell,r},d_{\ell,r}),\ldots,\mathbf{J}_{N_r}(\phi_{\ell,r},d_{\ell,r})\right)$, where each antenna gain from $n$-$th$ antenna is denoted as 
\begin{equation} \small
    \mathbf{J}_n(\phi_{\ell,r},d_{\ell,r}(n))=\begin{bmatrix}\mathcal{G}_{\text{C},n}(\phi_{\ell,r},d_{\ell,r}(n))&&\mathcal{G}_{\text{X},n}(\phi_{\ell,r},d_{\ell,r}(n))\\\mathcal{G}_{\text{X},n}(\phi_{\ell,r},d_{\ell,r}(n))&&-\mathcal{G}_{\text{C},n}(\phi_{\ell,r},d_{\ell,r}(n))\end{bmatrix}.
\end{equation}
Moreover, each element in the $n$-$th$ antenna gain is described as co-pol gain $\mathcal{G}_{\text{C},n}(\phi_{\ell,r},d_{\ell,r}(n))$ and cross-pol gain $\mathcal{G}_{\text{X},n}(\phi_{\ell,r},d_{\ell,r}(n))$. The \ac{NF} co-pol radiation pattern is
\begin{equation}
    \mathcal{G}_{\text{C},n}(\phi_{\ell,r},d_{\ell,r}(n))=\mathcal{G}_{\max}~f_{\mathrm{co}}(\phi_{\ell,r})\cdot\frac{1}{d_{\ell,r}(n)}\cdot e^{-j\frac{2\pi}{\lambda}d_{\ell,r}(n)},
\end{equation}
where $\mathcal{G}_{\mathrm{max}}$ is the maximum co-pol gain of the antenna, $f_{\mathrm{co}}(\phi_{\ell,r})$ is the normalized co-pol radiation pattern as a function of the angle $\phi_{\ell,r}$, (e.g., $\sin(\phi)$ for a dipole antenna, $\cos^k(\phi)$ for a patch antenna). The $e^{- j\frac {2\pi }\lambda d_{\ell,r}(n) }$ represents the phase shift due to the varying distance $d_{\ell,r}(n)$ between each antenna element and the scatterer.
\par Similarly, the cross-pol gain $\mathcal{G}_{X,n}(\phi_{\ell,r},d_{\ell,r}(n))$ in terms of
$\mathcal{G}_{C,n}(\phi_{\ell,r},d_{\ell,r}(n))$ and \ac{XPD} is represented as follows
\begin{equation}
    \mathcal{G}_{X,n}(\phi_{\ell,r},d_{\ell,r}(n))=\frac{\mathcal{G}_{C,n}(\phi_{\ell,r},d_{\ell,r}(n))}{\sqrt{\mathrm{XPD}_n}},
\end{equation}
where the \ac{XPD} which is described as the ratio of the co-pol signal's average received power to the cross-pol signal's average received power at each antenna element, causing a power imbalance between both polarization components
\begin{equation}\label{XPDFinal}
    \mathrm{XPD}_n =\frac{P_C (n)}{P_X (n)}= \left(\frac{\mathcal{G}_{C,n}(\phi_{\ell,r},d_{\ell,r}(n))}{\mathcal{G}_{X,n}(\phi_{\ell,r},d_{\ell,r}(n))}\right)^2 .
\end{equation}
\subsubsection{Visibility Region} In \ac{XL}-\ac{MIMO} systems, the concept of \acp{VR} is crucial due to the large aperture of the antenna array, where each antenna element might have different visibility to the surrounding environment, especially in the NF regime. The visibility determines which antennas contribute to the received signal for each multipath component. Each $\ell$-$th$ path might only be visible to a subset of antenna elements. This can be modeled using the visibility function $u_\ell$ defined as
\begin{equation}
u_{\ell}(n) =
\begin{cases} 
1, & n \in \mathcal{V}_{\ell} \\
0, & n \notin \mathcal{V}_{\ell}
\end{cases}~,
\end{equation}
where $\mathcal{V}_{\ell}$ represents the set of antennas within the VR of the $\ell$-$th$ path. This means that only antennas with $u_{\ell}(n) = 1$ will contribute to the received signal for the $\ell$-$th$ path.
\subsubsection{Depolarization Matrix}
The polarization of the signal can be affected by depolarization, leading to energy leakage between orthogonal polarization components (H-to-V or V-to-H). To model this, we introduce the depolarization matrix $\mathbf{X}_{\ell}$ for the $\ell$-$th$ path, modeled using a simple correlation model~as
\begin{equation}
\mathbf{X}_{\ell} = \sqrt{\frac{1}{1+\chi}} 
\begin{bmatrix}
e^{j\alpha_{\ell}^{HH}} & \sqrt{\chi} e^{j\alpha_{\ell}^{HV}} \\
\sqrt{\chi} e^{j\alpha_{\ell}^{VH}} & e^{j\alpha_{\ell}^{VV}}
\end{bmatrix},
\end{equation}
where $\chi$ is the inverse of \ac{XPD}, $\alpha_{\ell}^{\mathrm{XY}}$ is the phase change induced when going from X to Y polarization. Normalization is applied so that changing $\chi$ does not affect the depolarization matrix norm. The $\mathbf{X}_{\ell}$ matrix models how much of the signal's energy remains in its original polarization (co-pol) and how much leaks into the orthogonal polarization (cross-pol). The parameter $\chi$ controls the strength of this leakage.
\subsubsection{Rotation Matrix}
\begin{figure}
\centering 
\resizebox{1\columnwidth}{!}{
\includegraphics{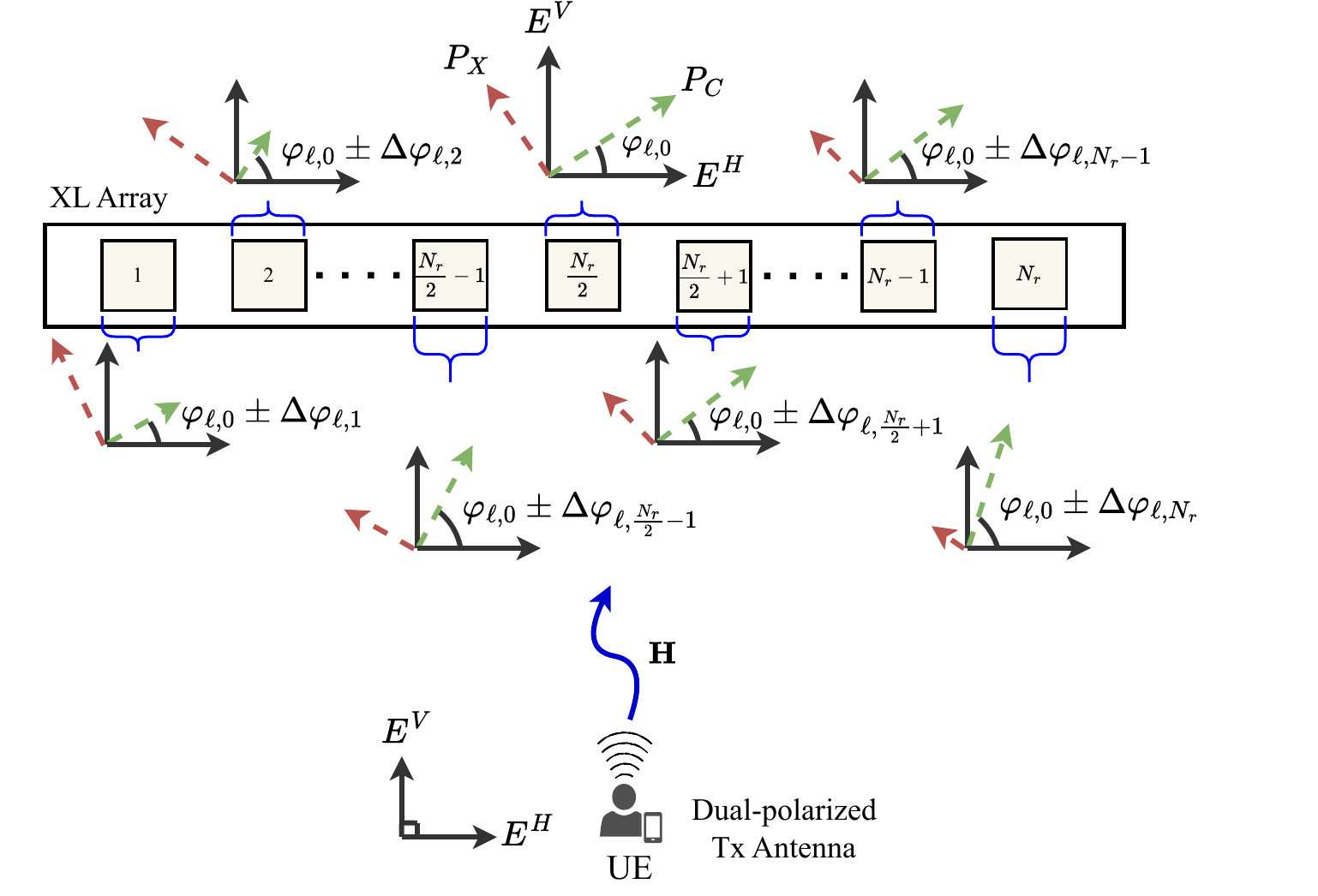}}
\caption{Antenna model with co-pol and cross-pol components at each antenna element at the XL-\ac{MIMO} array.}
\label{dpz}
\end{figure}
As illustrated in Fig. \ref{dpz}, the V and H gain patterns are related to the cross-pol and co-pol through a coordinate rotation where $\varphi$ is the polarization angle and $\mathbf{Q}_{\ell}(\varphi_{\ell}) \in \mathbb{C}^{2 \times 2}$ is the rotation matrix for the $\ell$-$th$ path
\begin{equation} \label{rotation}
\mathbf{Q}_{\ell}(\phi_{\ell}) = 
\begin{bmatrix}
\cos(\varphi_{\ell}) & \sin(\varphi_{\ell}) \\
-\sin(\varphi_{\ell}) & \cos(\varphi_{\ell})
\end{bmatrix}.
\end{equation}
The matrix rotates the coordinates by $\varphi_{\ell}$ in a 2D plane (x,~y). This transforms the local coordinates of the antenna elements or the received signal into a globally rotated coordinate system. It allows us to account for the fact that the signals may arrive at the antennas from different polarization angles.
\subsubsection{XL-\ac{MIMO}-based Polarization Shifts} Specifically in the case of \ac{NF} \ac{XL}-\ac{MIMO}, each antenna element in the array receives the signal with a slightly different polarization angle due to the geometric phase differences that arise from its position. The result is a polarization gradient across the array, as each element effectively sees the signal at a slightly different orientation. This relation is mathematically represented as
\begin{figure}[t]
	\centering
    \includegraphics[scale=0.55]{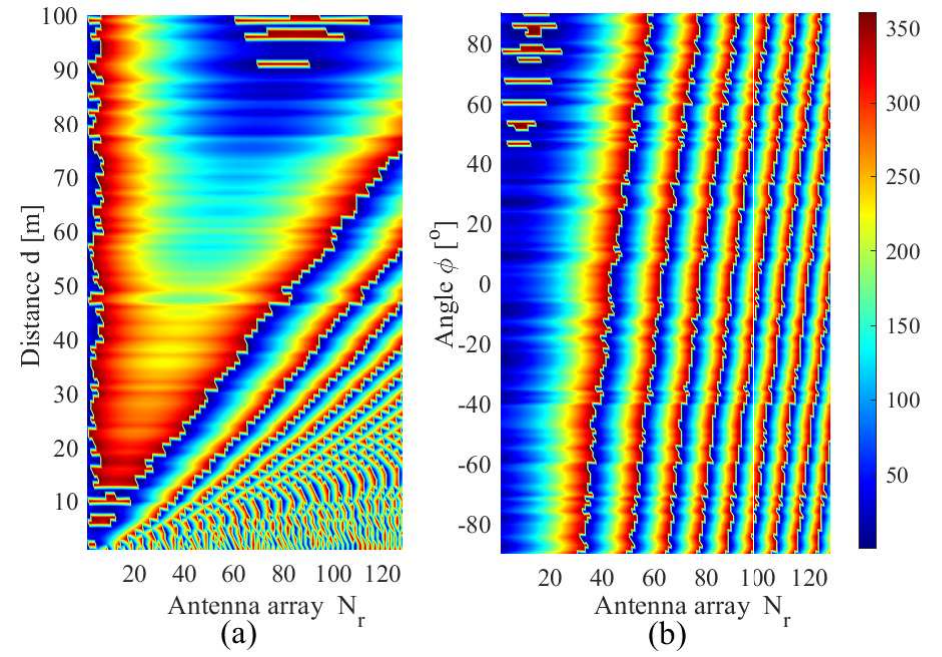}
	\caption{ Polarization angle distribution in XL-\ac{MIMO} systems for: (a) variation of polarization angle with distance $d$ and $N_r$, (b) variation of polarization angle with angle $\phi$ and $N_r$. The color bar represents the polarization angle in degrees $[^o]$.}
	\label{fig:PolarizationAnten}
\end{figure}
\begin{equation}
    \varphi_{\ell,n} = \varphi_{\ell,0} + \frac{2\pi}{\lambda} \Delta d_{\ell,n},
\end{equation}
where $\Delta d_{\ell,n} = \sqrt{\bar{d}_{\ell,r}^2 + (n d_\text{array})^2} - \bar{d}_{\ell,r}$, $\varphi_{\ell,0}$ is the initial polarization angle at the middle antenna element in the array from path $\ell$. Fig. \ref{fig:PolarizationAnten} illustrates this spatial variation of polarization angle across an \ac{XL}-\ac{MIMO} array as a function of (a) distance $d$ and (b) angle $\phi$. In the \ac{NF} regime, the curvature of the incident wavefront induces non-stationary polarization characteristics across the array, unlike the uniform polarization response observed in \ac{FF} \ac{MIMO}. Fig. \ref{fig:PolarizationAnten} (a) shows that at shorter distances ($<$30), the polarization angle exhibits rapid fluctuations due to spherical wavefront effects, while at greater distances, the variations become more structured. Fig. \ref{fig:PolarizationAnten} (b) highlights periodic polarization shifts with respect to $\phi$, which are linked to spatial phase differences across the array. These effects play a critical role in \ac{NF} radar and \ac{JRC} systems, where polarization diversity can be leveraged for interference suppression and target discrimination.
\subsection{Antenna Selection}
To mitigate \ac{XPD} and optimize sensing-communication coexistence, we introduce a dynamic polarization-aware \ac{AS} mechanism that selects a subset of antennas based on polarization alignment, interference minimization, and spatial efficiency. With the dynamic selection applied, not all antennas should be active, only a subset $\mathcal{S}$ of $N_r$ selected antennas contributes to the received signal. To extract only the selected antennas, we define the \ac{AS} matrix $\mathbf{S}$ as a binary diagonal matrix that selects the desired antennas. If the $n$-$th$ antenna is selected, then $S_{nn}=1$, otherwise $\mathbf{S}_{nn}=0$. The received signal in \eqref{yyy} after antenna selection becomes $\mathbf{\hat{y}} = \mathbf{S} \mathbf{y}(t).$ This selection allows us to mathematically integrate our proposed polarization-aware selection mechanism into the received signal model. The \ac{SINR} at the receiver is then given as
\begin{equation}
    \text{SINR} = \frac{|| \mathbf{S} \mathbf{H}_{C}\mathbf{E}||^2}{|| \mathbf{S} \mathbf{H}_{X}\mathbf{E}||^2 + \sigma_c^2} ,
\end{equation}
where~$\mathbf{H}_C =
\begin{bmatrix}
\textbf{h}_{V_rV_t} & 0 \\
0 & \textbf{h}_{H_rH_t}
\end{bmatrix},$~and~$\mathbf{H}_X =
\begin{bmatrix}
0 & \textbf{h}_{V_r H_t} \\
\textbf{h}_{H_rV_t} & 0
\end{bmatrix}$.
\section{Problem Formulation}
\par In this section, we formulate an optimization problem for polarization-aware \ac{AS} in \ac{NF} \ac{XL}-\ac{MIMO} \ac{JRC} systems, aiming to maximize the \ac{SE} and radar \ac{$P_d$} while minimizing \ac{XPD}. The problem incorporates \ac{AS}, polarization-dependent interference mitigation, and computational efficiency constraints to ensure practical feasibility. A key challenge in \ac{DP} \ac{JRC} multiplexing is polarization leakage, where channel depolarization leads to \ac{XPD} between \ac{JRC} signals \cite{10458884}. This interference arises due to multipath propagation, scattering, and reflections, degrading both communication \ac{SE} and radar \ac{$P_d$}. The issue is amplified in \ac{XL}-\ac{MIMO} due to its large-scale array structure and inherent non-stationarity. Unlike conventional \ac{MIMO}, where all antennas experience similar channel conditions, \ac{NF} \ac{XL}-\ac{MIMO} introduces distinct \acp{VR}, leading to polarization distortions across different subarrays. The goal is to select antennas maximizing \ac{SE} for communication while ensuring high \ac{$P_d$} for sensing, subject to polarization and computational constraints
\subsection{Communication-based Antenna Selection} \label{ASCOMM}
For communication, the objective is to maximize \ac{SE} by selecting antennas that improve SINR while minimizing cross-pol interference. The optimization problem is formulated as
\begin{subequations}
\begin{align}
   \mathbf{P1:} \quad &\underset{\mathcal{S}_c}{\text{max}} \quad \sum_{n \in \mathcal{S}_c} \log_2 \left( 1 + \text{SINR}_n \right),\label{eq:P1_obj}\\
   \text{s.t.} \quad &\mathcal{S}_c \subseteq \mathcal{V}, \quad |\mathcal{S}_c| \geq S_c^{\min}, \quad |\mathcal{S}_c| \leq N_c,\label{eq:P1_con1}\\
   & \frac{P_X(n)}{P_C(n)} \leq \gamma_{\text{comm}}, \quad \forall n \in \mathcal{S}_c,\label{eq:P1_con2}
\end{align}
\end{subequations}
where \( \text{SINR}_n \) is the \ac{SINR} at the $n$-$th$ antenna, contributing to the overall \ac{SE}, \( \mathcal{S}_c \) represents the set of selected antennas for communication, \( S_c^{\min} \) is the minimum number of selected antennas, ensuring fairness when multiple VRs exist. Constraint \eqref{eq:P1_con1} enforces selection within the \ac{VR} while ensuring that the number of selected antennas; is at least the size of the smallest \ac{VR} across all paths, $S_c^{\min} = \min \left( |\mathcal{V}_{1}|, |\mathcal{V}_{2}|, ..., |\mathcal{V}_L| \right)$ does not exceed the maximum allowed antennas \( N_c \). Constraint \eqref{eq:P1_con2} enforces a cross-pol power ratio threshold, ensuring that antennas with excessive polarization leakage are not selected.
\par When only a single \ac{VR} is observed (\( L = 1 \)), the selection adapts based on system requirements
\begin{equation} \label{priority}
    S_c = \begin{cases}
    \frac{|\mathcal{V}_{1}|}{2}, & \text{(Fairness-based Selection)} \\
    \lambda |\mathcal{V}_{1}|, & \text{(Adaptive-based Selection, \(\lambda\))}
    \end{cases}
\end{equation}
where \( \lambda \) is an adjustable factor, \( \lambda > 0.5 \) prioritizes communication centric applications, and \( \lambda < 0.5 \) prioritizes sensing centric applications. The final \ac{AS} for communication is
\begin{equation}
    S_c = \bigcup_{\ell} S_c^{\ell}, \quad \forall \ell \in \{1, \dots, L\}
\end{equation}
where \( S_c^{\ell} \subseteq \mathcal{V}_{\ell} \) is \ac{AS} from each \ac{VR}. If multiple \acp{VR} exist, then ensure $|S_c| \geq S_c^{\min}$ for fairness across different \acp{VR}.
\subsection{Sensing-based Antenna Selection}
For radar sensing, the objective is to maximize detection probability by selecting antennas that provide strong co-pol radar returns. The problem is formulated as
\begin{subequations}
\begin{align}
    \mathbf{P2:} \quad &\underset{\mathcal{S}_s}{\text{max}} \quad P_d = Q\left( \frac{\tau - \sum_{n \in \mathcal{S}_s} P_n^{\text{radar}}}{\sigma_n} \right) \label{eq:P2_obj}\\
    \text{s.t.} \quad & \mathcal{S}_s \subseteq \mathcal{V}, \quad |\mathcal{S}_s| \geq S_s^{\min}, \quad |\mathcal{S}_s| \leq N_s\label{eq:P2_con1}\\
    & \frac{P_C(n)}{P_X(n)} \geq \gamma_{\text{radar}}, \quad \forall n \in \mathcal{S}_s\label{eq:P2_con2}
\end{align}
\end{subequations}
where \ac{$P_d$} is modeled using the Q-function, \( \tau \) is the detection threshold, and \( \sigma_n \) is the noise variance at the receiver. \( P_n^{\text{radar}} \) represents the received radar power at the \( n \)-$th$ antenna, \( \mathcal{S}_s \) is the set of selected antennas for sensing and \( S_s^{\min} \) is the minimum number of antennas required to ensure fairness in multi-\ac{VR} scenarios. Constraint \eqref{eq:P2_con1} enforces selection within the \ac{VR} while ensuring that the number of selected antennas is at least the size of the smallest VR across all paths, $S_s^{\min} = \min \left( |\mathcal{V}_{1}|, |\mathcal{V}_{2}|, ..., |\mathcal{V}_L| \right)$ and does not exceed the maximum allowed antennas \( N_s \). Constraint \eqref{eq:P2_con2} ensures co-pol dominance, meaning antennas where the co-pol radar power is significantly higher than the cross-pol power are prioritized. Similar to (\ref{priority}), here also the same concept is ensured for fairness-based and adaptive-based selection for sensing-centric and communication-centric applications.
\par The final set of selected sensing antennas is given by
\begin{equation}
    S_s = \bigcup_{\ell} S_s^{\ell}, \quad \forall \ell \in \{1, \dots, L\}
\end{equation}
where \( S_s^{\ell} \subseteq \mathcal{V}_{\ell} \) represents antennas selected from each VR. If multiple VRs exist, we ensure that $|S_s| \geq S_s^{\min}$ to guarantee fairness across different \acp{VR}.
\section{Proposed Antenna Selection Scheme}
In this section, we propose a Greedy-based \ac{AS} scheme for \ac{XL}-\ac{MIMO} \ac{JRC} systems. The objective is to identify the most effective subset of antennas for both sensing and communication, ensuring high SE, improved SINR, and enhanced radar detection probability while mitigating \ac{XPD}. At first, each antenna in the \ac{XL}-\ac{MIMO} array captures both polarization components (H and V), which undergo polarization shifts due to propagation effects. Furthermore, each antenna element introduces an additional random polarization deviation of \(\pm\Delta\), leading to further polarization misalignment. Additionally, depolarization effects cause power imbalance between the two polarizations \cite{guo2016advances}, influencing the \ac{AS} process.
\par To address these issues, we introduce a power-aware and polarization-aware selection approach based on the following
\begin{itemize}
    \item Received Co-Pol Power Levels: Evaluate whether the antenna is better suited for sensing or communication.
    \item \ac{XPD} Suppression: Ensures antennas with high polarization leakage are not selected.
    \item Spatial Consistency: Maintains smooth antenna selection transitions across the array, avoiding sudden variations.
\end{itemize}

Starting from the first criteria, the received power at the $n$-$th$ antenna for each polarization component is given by
\begin{equation} \label{CA1}
    P_H(n) = \mathbb{E}\left[\left| {y}_n^{H} \right|^2\right], \quad 
    P_V(n) = \mathbb{E}\left[\left| {y}_n^{V} \right|^2\right].
\end{equation}
where \( P_H(n) \) represents the received power of the H polarization component (co-pol OFDM) and \( P_V(n) \) represents the received power of the V polarization component (co-pol chirp). To quantify the power imbalance between them, we define the polarization power imbalance factor as
\begin{equation} \label{CA3}
    \Gamma_n = \frac{P_H(n)}{P_V(n)},
\end{equation}
where a high $\Gamma_n$ indicates that H-polarized signals are dominant, and a low $\Gamma_n$ indicates V-polarized dominance. The cross-pol power components at each antenna are given by
\begin{equation} \label{CA2}
    P_X^{H \to V}(n) = \mathbb{E}\left[\left| {y}_n^{V,H} \right|^2\right], \quad 
    P_X^{V \to H}(n) = \mathbb{E}\left[\left| {y}_n^{H,V} \right|^2\right],
\end{equation}
where \( P_X^{H \to V}(n) \) represents cross-pol power leakage from \ac{OFDM} (H) into the chirp (V) component and \( P_X^{V \to H}(n) \) represents cross-pol power leakage from chirp (V) into the OFDM (H) component. 
\begin{algorithm}
\caption{Proposed Greedy-based AS for XL-\ac{MIMO} JRC}
\label{algo:AS}
\begin{algorithmic}[1]
\STATE \textbf{Input:} \( \mathbf{y}_n^{H}, \mathbf{y}_n^{V} \), \( P_X^{H \to V}(n), P_X^{V \to H}(n) \), $\epsilon$
\STATE \textbf{Output:} \( \mathcal{S} \)
\STATE \textbf{Initialization:} Set \( \mathcal{S} = \emptyset \)
\FOR{$n= 1:N$}
    \STATE Compute received power for each polarization using \eqref{CA1}.
    \STATE Compute cross-pol power leakage using \eqref{CA2}.
    \STATE \textbf{Step 1: Antenna Selection}
    \IF{\( P_H(n) > P_X^{V \to H}(n) \)}
        \STATE Assign antenna \( n \) for \textbf{communication} (\( S_n = 1 \))
    \ELSIF{\( P_V(n) > P_X^{H \to V}(n) \)}
        \STATE Assign antenna \( n \) for \textbf{radar sensing} (\( S_n = 1 \))
    \ELSE
        \STATE Discard antenna \( n \) (\( S_n = 0 \))
    \ENDIF
    \STATE \textbf{Step 2: Spatial Consistency}
    \IF{\( |P_H(n) - P_V(n)| \leq \epsilon \)}
        \STATE Assign \( S_n = S_{n-1} \) (Maintain spatial coherence)
    \ENDIF
\ENDFOR
\STATE \textbf{Step 3: Final Selection}\\
$\mathcal{S} = \bigcup_{\ell} \mathcal{S}_{\ell}, \quad \forall \ell \in \{1, \dots, L\}.$
\STATE \textbf{Return} \( \mathcal{S} \)
\end{algorithmic}
\end{algorithm}
\par To determine whether an antenna should be assigned for communication or sensing, we use the following selection rule
\begin{equation} \label{CA4}
    S_n =
    \begin{cases}
        1, & \text{if } P_H(n) > P_X^{V \to H}(n), \quad \text{(AS for comm.)}, \\
        1, & \text{if } P_V(n) > P_X^{H \to V}(n), \quad \text{(AS for sensing)}, \\
        0, & \text{otherwise}.
    \end{cases}
\end{equation}
where if \( P_H(n) \) is greater than the cross-pol interference \( P_X^{V \to H}(n) \), then the antenna is assigned for communication, and if \( P_V(n) \) is greater than the cross-pol interference \( P_X^{H \to V}(n) \), then the antenna is assigned for radar sensing. If the power levels are nearly equal, an additional constraint is applied to maintain spatial consistency.

In scenarios where the power levels of both polarizations are nearly equal, selection is smoothed by ensuring antennas in the same VR follow a consistent trend
\begin{equation}
    S_n =
    \begin{cases}
        S_{n-1}, & \text{if } |P_H(n) - P_V(n)| \leq \epsilon, \\
        S_n, & \text{otherwise}.
    \end{cases}
\end{equation}
where \( \epsilon \) is a small threshold to detect near-equal power levels. The \ac{AS} \( S_n \) follows its previous antenna’s selection \( S_{n-1} \), ensuring spatial smoothness.

The final set of selected antennas across all VRs is
\begin{equation}
    \mathcal{S} = \bigcup_{\ell} \mathcal{S}_{\ell}, \quad \forall \ell \in \{1, \dots, L\}.
\end{equation}
where \( \mathcal{S} \in \{S_c, S_s\}\) is the final set of selected antennas, and \( \mathcal{S}_{\ell} \) is the subset of antennas within the \( \ell \)-th VR. The proposed AS scheme is summarized in Algorithm \ref{algo:AS}. \\
\noindent \textbf{Polarization-induced interference mitigation}: From \eqref{depol}, the cross-pol interference components affect the received signal. To suppress this, we adopt the depolarization mitigation technique presented in \cite{10458884}, leading to a refined received signal model
\begin{equation}\label{depol2}
\begin{aligned}
y^{\text{V}}_n(t) = h^n_{V_r V_t} E^V(t) + w^V_n(t) ,\\
y^{\text{H}}_n(t) = h^n_{H_r H_t} E^H(t) + w^H_n(t) ,
\end{aligned}
\end{equation}
where interference components are significantly suppressed, although some residual interference may persist. For both $y^{\text{V}}_n(t)$ and $y^{\text{H}}_n(t)$ conventional radar and communication signal processing is applied as in \cite{10458884} and \cite{zegrar2022common}.
\subsection{Complexity Analysis} \label{Compi}
\par The proposed \ac{AS} scheme for \ac{XL}-\ac{MIMO} operates in a \ac{NF} environment. The number of paths $L$, observed by a~\ac{UE} determines the number of \acp{VR} available for \ac{AS}. The computational complexity of the proposed \ac{AS} is analyzed as follows:
\begin{enumerate}
\item \textit{Computing the Received Power Components:} For each antenna \( n \), we compute the received power for both co-pol and cross-pol components using (\ref{CA1}) and (\ref{CA2}). Since these computations must be performed for all \( N \) antennas, the complexity is \( \mathcal{O}(N) \).
\item \textit{Computing the Polarization Power Imbalance Factor:} The polarization power imbalance factor at each antenna is computed from (\ref{CA3}). This step involves a division operation per antenna, yielding a complexity of \( \mathcal{O}(N) \).
\item \textit{Antenna Selection Based on Polarization:} The selection criteria involve threshold-based comparisons from (\ref{CA4}). Each decision requires a constant-time operation per antenna, leading to a total complexity of \( \mathcal{O}(N) \).
\item \textit{Visibility Region Consideration:} Since each \ac{UE} perceives multiple \acp{VR}, fairness in \ac{AS} requires selecting a minimum number of antennas across \( L \) VRs. Finding the minimum from \( L \) sets of antennas (each containing at most \( N \) elements) requires \( \mathcal{O}(L N) \).
\item \textit{Sorting for Fairness and Load Balancing:} To ensure fairness, antennas within each \ac{VR} are sorted based on received power, requiring \( \mathcal{O}(N \log N) \) per VR. Since sorting is performed across \( L \) \acp{VR}, the total sorting complexity is \( \mathcal{O}(L N \log N) \).
\item \textit{Final Complexity Expression:} Summing up the complexity of all operations, we obtain $
\mathcal{O}(N) + \mathcal{O}(N) + \mathcal{O}(N) + \mathcal{O}(L N) + \mathcal{O}(L N \log N)$. Since the sorting term dominates, the overall complexity of the proposed method is, $\mathcal{O}(L N \log N)$. For multiple \acp{UE} the complexity becomes, $\mathcal{O}(KL N \log N)$.
\end{enumerate}
\par The proposed \ac{AS} scheme is the first in the literature to jointly incorporate both \ac{JRC} and polarization effects in \ac{XL}-\ac{MIMO}. Due to the distinctiveness of this approach, direct comparisons with existing works are not feasible. However, to establish a benchmark, we compare the computational complexity of our method with existing \ac{AS} schemes in \ac{XL}-\ac{MIMO} that focus exclusively on communication \cite{marinello2020antenna, de2021quasi}, as radar sensing has not been considered in these prior works.
\par The complexity in \cite{marinello2020antenna} is based on four different schemes. Their complexities are; highest received normalized power scheme (HRNP), $\mathcal{O}(MK + M \log M)$ where \( M \) is number of antennas and \( K \) is number of \acp{UE}, local search (LS)-based \ac{AS} $\mathcal{O}(N_{\text{it}} MK^2)$, where \( N_{\text{it}} \) is the number of iterations, genetic algorithm (GA) scheme, $ \mathcal{O}(N_{\text{it}} p_{\text{GA}} MK^2)$, where \( p_{\text{GA}} \) is the population size of the GA, and particle swarm optimization (PSO), $\mathcal{O}(N_{\text{it}} p_{\text{PSO}} MK^2)$, where \( p_{\text{PSO}} \) is number of particles in PSO. Similarly, GA scheme in \cite{de2021quasi} has complexity of $\mathcal{O}(K^3 T(N_p + N_e) + K^3 N_e + K^2 T N (N_p + N_e) + K^2 N_e)$, where \( T \) is number of generations, \( N_p \) is population size, and \( N_e \) is number of elite individuals. In contrast, our proposed \ac{AS} has complexity of $\mathcal{O}(K L N \log N)$. Unlike GA and PSO, which require $\mathcal{O}(K^2)$ to $\mathcal{O}(K^3)$ operations due to iterative evaluations, our method efficiently selects antennas without costly population-based optimization. The GA-based scheme in \cite{de2021quasi} scales as $\mathcal{O}(K^3)$, making it computationally intensive for large-scale systems. Similarly, GA and PSO in \cite{marinello2020antenna} require $\mathcal{O}(K^2)$ operations, leading to scalability challenges. In contrast, our method’s $\mathcal{O}(K L_u N \log N)$ complexity ensures better scalability for XL-\ac{MIMO}, reducing processing time while maintaining fairness and mitigating cross-pol interference.
\section{System Analysis}
This section analyzes the proposed \ac{AS} scheme in terms of \ac{SER} for communication and \ac{$P_d$} for sensing under varying \ac{SNR}, polarization shifts, and interference conditions.
\subsection{Symbol Error Rate}
\par In this section, we derive the \ac{SER} performance of the proposed \ac{AS} scheme while considering both polarization-induced interference mitigation and \ac{AS}. The objective is to evaluate how proposed \ac{AS} improves the received \ac{SNR} and reduces the \ac{SER} in \ac{XL}-\ac{MIMO} \ac{JRC} systems. The total power of the useful signal received at the radar receiver in the presence of \ac{XPD} is given as \cite{6779686}
\begin{equation} \label{eq:received_power_xpd}
    P'_R = \frac{P_T}{1+\chi},
\end{equation}
where \( P_T \) consists of co-pol (\( P_T^C \)) and cross-pol (\( P_T^X \)) components as \( P_T = P_T^C + P_T^X \). The \( \chi \) models cross-pol leakage, we have \( P_T^X = \chi P_T^C \), leading to the co-pol component, $P_T^C = {P_T}/{1+\chi}.$ Since only the co-pol signal is useful, the effective received power is $P'_R = P_T^C = {P_T}/{1+\chi}.$ Moreover, \( \chi \) is within \( 0 \leq \chi \leq 1 \), a higher \( \chi \) implies greater leakage and degraded signal reception. The received \ac{SNR} without mitigation is
\begin{equation} \label{eq:snr_without_mitigation}
    \text{SNR}_r = \frac{P'_R}{N_0} = \frac{P_T}{(1+\chi)N_0},
\end{equation}
where $N_0$ is the power spectral density of Gaussian noise.
\par If polarization mitigation is applied, the received signal components are refined, and the effective power is given~as
$P^{H''} = \mathbb{E}[|E^H|^2],$ and $ P^{V''} = \mathbb{E}[|E^V|^2]$. The \ac{SNR} after polarization mitigation improves to
\begin{equation} \label{eq:snr_with_mitigation}
    \text{SNR}'_r = \frac{P^{H''} + P^{V''}}{N_0} = \frac{P_T}{N_0}.
\end{equation}
\par With the proposed \ac{AS}, antennas with high interference are excluded, retaining only those with strong signals. The effective power across selected antennas is $P_R^{\text{AS}} = \sum_{n \in \mathcal{S}_c} P_n^{\text{useful}},$ where \( P_n^{\text{useful}} \) is the received power at the \( n \)-$th$ selected antenna. The corresponding \ac{SNR} with \ac{AS} is $ \text{SNR}_{\text{AS}} = {P_R^{\text{AS}}}/{N_0}.$ Furthermore, by excluding antennas that suffer from high \ac{XPD}, the effective \ac{XPD} factor is reduced
\begin{equation} \label{eq:xpd_after_as}
\eta_{\text{AS}}(\chi) = \frac{1}{1 + \chi_{\text{AS}}},
\end{equation}
where \( \chi_{\text{AS}} \) represents the adjusted \ac{XPD} factor after selecting the most favorable antennas, ensuring that \( \chi_{\text{AS}} < \chi \). The \ac{SER} in a fading channel is given by the integral
\begin{equation} \label{eq:ser_general}
\text{SER} = \int_{0}^{\infty} Q\left(\sqrt{2 \frac{P_T}{(1+\chi) N_0} \gamma} \right) f_{\gamma}(\gamma) d\gamma.
\end{equation}
where \( \gamma \) is the instantaneous \ac{SNR}, and \( f_{\gamma}(\gamma) \) is the PDF of \( \gamma \). For a Rayleigh fading channel, \( \gamma \) follows an exponential distribution $f_{\gamma}(\gamma) = \frac{1}{\bar{\gamma}} e^{-\gamma/\bar{\gamma}},$ where \( \bar{\gamma} \) is the average \ac{SNR}, substituting this into (\ref{eq:ser_general}) yields
\begin{equation} \label{eq:ser_integral}
\text{SER} = \int_{0}^{\infty} Q\left(\sqrt{2 \frac{P_T}{(1+\chi) N_0} \gamma} \right) \frac{1}{\bar{\gamma}} e^{-\gamma/\bar{\gamma}} d\gamma.
\end{equation}
Using Craig’s formula for the Q-function \cite{9058698}, $Q(x) = \frac{1}{\pi} \int_{0}^{\frac{\pi}{2}} \exp\left(-\frac{x^2}{2 \sin^2 \theta}\right) d\theta$, and solving using contour integration, the \ac{SER} without \ac{AS} and without mitigation is
\begin{equation} \label{eq:ser_without_as}
\text{SER} = \frac{1}{2} \left(1 - \sqrt{\frac{\bar{\gamma}}{1 + \bar{\gamma}}} \right).
\end{equation}
\par When \ac{AS} is incorporated, the effective SNR is improved, leading to a modified SER expression
\begin{equation} \label{eq:ser_with_as}
\text{SER}_{\text{AS}} = \frac{1}{2} \left(1 - \sqrt{\frac{\bar{\gamma}_{\text{AS}} \eta_{\text{AS}}(\chi)}{1 + \bar{\gamma}_{\text{AS}} \eta_{\text{AS}}(\chi)}} \right).
\end{equation}
\par The impact of \ac{AS} and polarization mitigation on \ac{SER} improvement can be expressed as
\begin{equation} \label{eq:ser_gain} \small
    \Delta \text{SER} = \text{SER} - \text{SER}_{\text{AS}} = \frac{1}{2} \left( \sqrt{\frac{\bar{\gamma}}{1+\bar{\gamma}}} - \sqrt{\frac{\bar{\gamma}_{\text{AS}} \eta_{\text{AS}}(\chi)}{1+\bar{\gamma}_{\text{AS}} \eta_{\text{AS}}(\chi)}} \right).
\end{equation}
Since cross-pol leakage \( \chi \) degrades the effective \ac{SNR}, its impact on SER is $\bar{\gamma}_{\text{eff}} ={\bar{\gamma}}/{1+\chi}.$ At high \ac{SNR} (\( \bar{\gamma} \gg 1 \)), SER approaches zero, $\lim_{\bar{\gamma} \to \infty} \text{SER} \approx 0$. At low \ac{SNR} (\( \bar{\gamma} \ll 1 \)), \ac{XPD} causes severe degradation, $\lim_{\bar{\gamma} \to 0} \text{SER} \approx \frac{1}{2}.$ Polarization mitigation and \ac{AS} reduce SER by enhancing received power and suppressing interference, making \ac{AS} essential for minimizing SER degradation in DP XL-\ac{MIMO} \ac{JRC} systems, especially under high-\ac{XPD} conditions.
\subsection{Probability of detection}
Once the \ac{AS} for radar receivers is allocated, radar signal processing is performed. To mathematically formulate radar target detection, we consider the received chirp signal ${E}^{V}(t)$ after analog mixing. The goal is to distinguish between two hypotheses in a binary hypothesis test $\mathcal{H}_1$, where a target is present, and $\mathcal{H}_0$, where only noise is present. The received signal model is expressed as
\begin{equation}
\begin{aligned}
        &\mathcal{H}_1: Y = {E}^{V}(t) + w^{V}(t),\\ 
        &\mathcal{H}_0: Y = w^{V}(t),
\end{aligned}
\end{equation}
where $w^{V}(t)$ represents the additive noise. This forms a standard binary hypothesis testing problem, where a \textit{linear detector} is applied \cite{10012421}, $z = |Y| \underset{\mathcal{H}_0}{\overset{\mathcal{H}_1}{\stackrel{<}{>}}} \mathcal{T},$ where $\mathcal{T}$ is the detection threshold. Under $\mathcal{H}_0$, the amplitude $z$ follows a Rayleigh distribution
\begin{equation}
    p_z\left(z \mid \mathcal{H}_0\right)= 
    \begin{cases}
    \frac{2 z}{\sigma_{\varphi}^2} \exp \left(-\frac{z^2}{\sigma_{\varphi}^2}\right), & z \geq 0 \\ 
    0, & z<0.
    \end{cases}
\end{equation}
The \ac{$P_{fa}$} is obtained by integrating the Rayleigh-distributed noise over the threshold
\begin{equation}
    P_{fa}=\int_{\mathcal{T}}^{\infty} p_z\left(z \mid \mathcal{H}_0\right) \mathrm{d} z = \exp \left(-\frac{\mathcal{T}^2}{\sigma_{\varphi}^2}\right).
\end{equation}
Rearranging for $\mathcal{T}$ at a given \ac{$P_{fa}$}, $ \mathcal{T}=\sigma_{\varphi} \sqrt{-\ln P_{fa}}.$ Here, $\sigma_{\varphi}$ represents the combined effects of noise and residual interference. Since detection is performed in each delay-Doppler bin, $\mathcal{T}$ is adjusted dynamically to maintain a constant $P_{fa}$ \cite{delamou2024interference}. Under hypothesis $\mathcal{H}_1$, the amplitude $z$ follows a Rician distribution
\begin{equation}
    p_z\left(z \mid \mathcal{H}_1\right) 
    = \begin{cases}
    \frac{2 z}{\sigma_{\varphi}^2} \exp \left[-\frac{\left(z^2+E^2\right)}{\sigma_{\varphi}^2}\right] 
    I_0\left(\frac{2 E z}{\sigma_{\varphi}}\right), & z \geq 0, \\
    0, & z<0.
    \end{cases}
\end{equation}
The \ac{$P_d$} is then given by $P_d=\int_{\mathcal{T}}^{\infty} p_z\left(z \mid \mathcal{H}_1\right) \mathrm{d} z$. By substituting the threshold $\mathcal{T}$, the closed-form expression for \ac{$P_d$} is derived as
\begin{equation}
    P_d = Q_1\left(\sqrt{\frac{2 E^2}{\sigma_{\varphi}^2}}, \sqrt{\frac{2 \mathcal{T}^2}{\sigma_{\varphi}^2}}\right),
\end{equation}
which, after substituting $\mathcal{T}$ in terms of \ac{$P_{fa}$}, simplifies to
\begin{equation}
    P_d = Q_1\left(\sqrt{\frac{2 |E|^2}{\mathbb{E}\left[|\varphi|^2\right]}}, \sqrt{-2 \ln P_{fa}}\right).
\end{equation}
Since the \ac{SNR} is defined as $\mathrm{SINR} = \frac{|E|^2}{\mathbb{E}\left[|\varphi|^2\right]},$ the final expression for $P_d$ is
\begin{equation} \label{pdderi}
    P_d= Q_1\left(\sqrt{2 \cdot \mathrm{SINR}}, \sqrt{-2 \ln P_{fa}}\right),
\end{equation}
where $Q_1(a, b)$ is the first-order Marcum $Q$-function. To achieve a specific \ac{$P_d$} at a given \ac{$P_{fa}$}, the required SNR is
\begin{equation}
   \mathrm{SNR} = \frac{\left(\frac{P_d}{P_{fa}}\right)^{1 / R_r}-1}{1-P_d^{1 / R_r}},
\end{equation}
where $R_r$ is the number of reference cells in the detection window.
\par The proposed AS scheme improves radar detection by selecting antennas that maximize the received signal quality while mitigating interference. Given that radar detection relies on the received (\ac{SNR}), the effective post-AS SNR is
\begin{equation} \label{eq:SNR_AS}
    \mathrm{SNR}_{\text{AS}} = \frac{\sum_{n \in \mathcal{S}_s} P_n^{\text{useful}}}{\mathbb{E}\left[|\varphi|^2\right]},
\end{equation}
where \( P_n^{\text{useful}} \) is the received power at each selected antenna. The AS reduces interference, enhancing effective SNR. The \ac{$P_d$} depends on \ac{SNR} as in \eqref{pdderi}. With AS, the improved \ac{$P_d$} is $ P_d^{\text{AS}} = Q_1\left(\sqrt{2 \cdot \mathrm{SNR}_{\text{AS}}}, \sqrt{-2 \ln P_{fa}}\right)$.
\par The \ac{$P_d$} gain from AS is $\Delta P_d = P_d^{\text{AS}} - P_d$. While the total received power is, $P_R = \sum_{n=1}^{N} \left( P_n^{\text{useful}} + P_n^{\text{interf}} \right).$ Since AS retains antennas with minimal interference, post-selection power is denoted as, $P_R^{\text{AS}} = \sum_{n \in \mathcal{S}_s} P_n^{\text{useful}}.$ This leads to SNR improvement due to AS as
\begin{equation} \label{eq:snr_improvement}
    \frac{\mathrm{SNR}_{\text{AS}}}{\mathrm{SNR}} = \frac{\sum_{n \in \mathcal{S}_s} P_n^{\text{useful}}}{\sum_{n=1}^{N} \left( P_n^{\text{useful}} + P_n^{\text{interf}} \right)}.
\end{equation}
\begin{figure}[ht]
\centering  
\subfloat[]{
  \includegraphics[width=41.9mm,height=35mm]{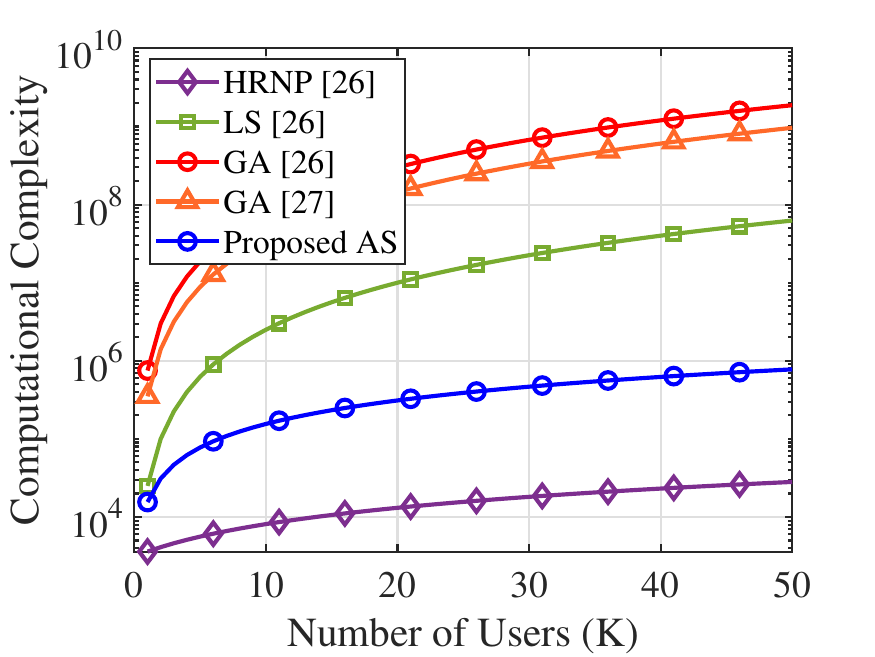}
}
\subfloat[]{
  \includegraphics[width=41.9mm,height=35mm]{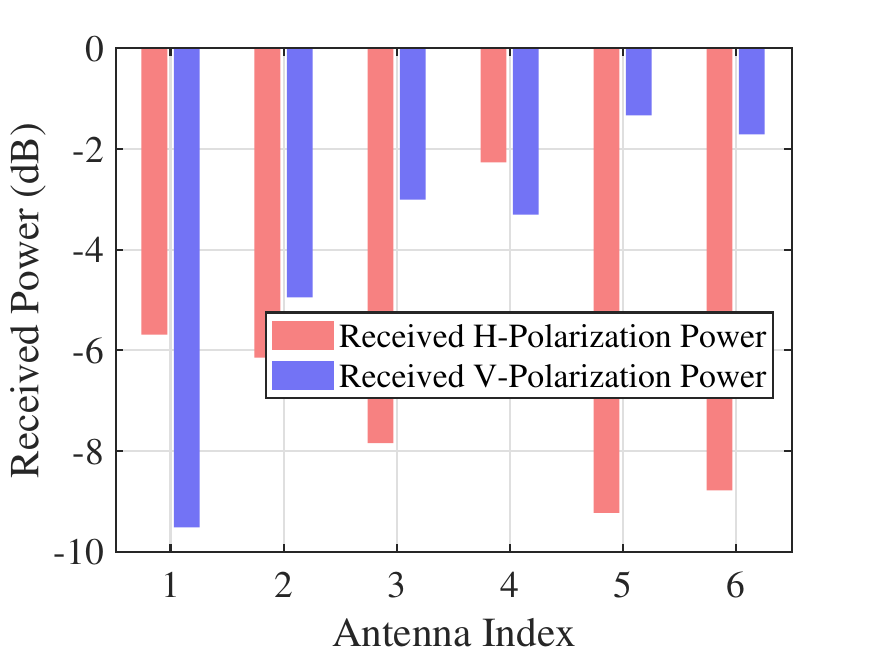}
  }
\caption{(a) Complexity analysis comparison of proposed \ac{AS} with \cite{marinello2020antenna, de2021quasi}, and (b) the power imbalance between the co- and cross-pol components (H/V).}
\label{COMPLEXsIMU}
\end{figure}
The AS removes antennas with interference,
$\sum_{n \in \mathcal{S}_r} P_n^{\text{interf}} \ll \sum_{n=1}^{N} P_n^{\text{interf}}$. This leads to a higher effective SINR
\begin{equation} \label{eq:sinr_with_as}
    \mathrm{SINR}_{\text{AS}} = \frac{\sum_{n \in \mathcal{S}_s} P_n^{\text{useful}}}{\sum_{n \in \mathcal{S}_s} P_n^{\text{interf}} + \mathbb{E}\left[|\varphi|^2\right]}.
\end{equation}
Since \ac{SINR} directly impacts detection, the \ac{$P_d$} with AS is given by $  P_d^{\text{AS}} = Q_1\left(\sqrt{2 \cdot \mathrm{SINR}_{\text{AS}}}, \sqrt{-2 \ln P_{fa}}\right).$ The final expression for SNR improvement due to AS is derived by normalizing the post-AS power levels
\begin{equation} \label{eq:snr_as_final}
    \mathrm{SNR}_{\text{AS}} = \frac{\sum_{n \in \mathcal{S}_s} P_n^{\text{useful}}}{\sum_{n=1}^{N} P_n^{\text{useful}}} \cdot \mathrm{SNR}.
\end{equation}
The proposed \ac{AS} significantly enhances the \ac{$P_d$} by increasing the effective received power and suppressing the interference. 

\section{Simulation Results}
\par This section evaluates the proposed \ac{AS} using metrics such as computational complexity, \ac{SE}, \ac{SINR}, and \ac{$P_d$}. The simulations are conducted for an XL-MIMO system with a uniform linear array with $N = 256$, and a carrier frequency of $f_c = 28$ GHz. The system models a propagation environment where each \ac{VR} exhibits polarization-dependent variations, with polarization angular shifts from $[0, 2\pi]$, $\mathcal{G}_{\mathrm{max}} = 1$, and $\alpha_{\ell}^{XY}\sim\mathcal{U}(0, 2\pi)$. The received co-pol and cross-pol components experience \ac{XPD} of 10 dB and $\chi=0.1$ dB. The \ac{$P_d$} is analyzed across \ac{SNR} values from $-10$ dB to $30$ dB.
\vspace{-3mm}
\subsection{Complexity Analysis}
This section evaluates the proposed \ac{AS} scheme's complexity and scalability compared to HRNP, LS, and GA-based methods. Fig. \ref{COMPLEXsIMU}(a) demonstrates that GA-based schemes exhibit cubic complexity, $\mathcal{O}(K^3)$ , due to iterative updates, making them infeasible for XL-MIMO. In contrast, the proposed AS leverages \ac{VR}-based selection, significantly reducing computational overhead while ensuring efficient antenna selection. HRNP, though computationally simple, suffers from poor selection accuracy, while LS improves performance at the cost of higher overhead. The proposed AS maintains controlled complexity, making it ideal for next-generation JRC applications where real-time scalability is critical. Fig. \ref{COMPLEXsIMU}(b) illustrates the polarization-dependent power imbalance across antennas, a key factor in XL-MIMO. Due to channel-induced mismatches and depolarization, received power varies for H (red) and V (blue) polarization components. The proposed AS dynamically selects antennas that maximize signal reception while suppressing and mitigating leakage.
\begin{figure}
\centering 
\resizebox{0.65\columnwidth}{!}{
\includegraphics{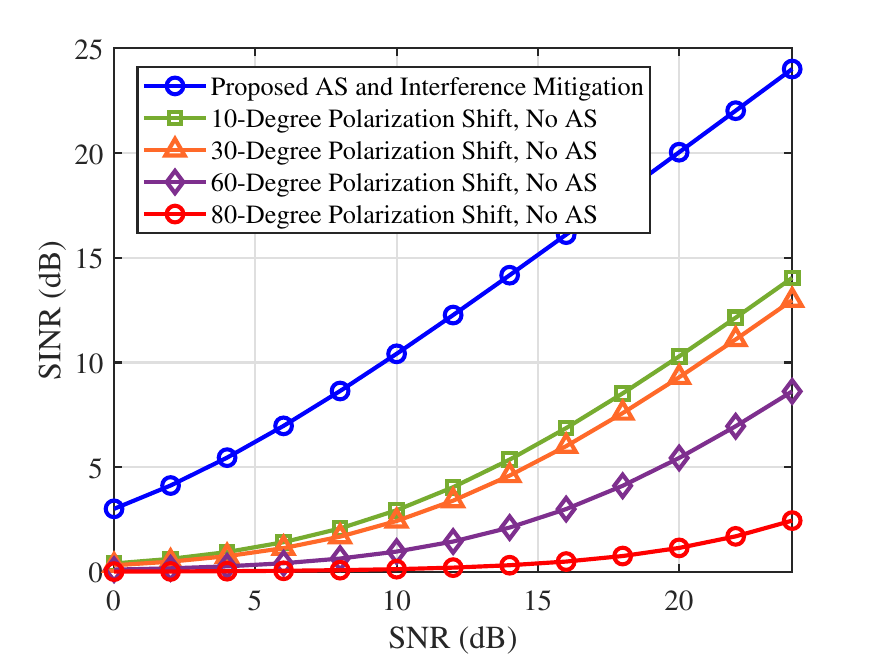}}
\caption{\ac{SINR} vs \ac{SNR} for different polarization angular shifts.}
\label{sinrComn}
\end{figure}
\vspace{-3mm}
\subsection{Communication Performance}
\begin{figure}
\centering  
\subfloat[]{
  \includegraphics[width=41.9mm,height=35mm]{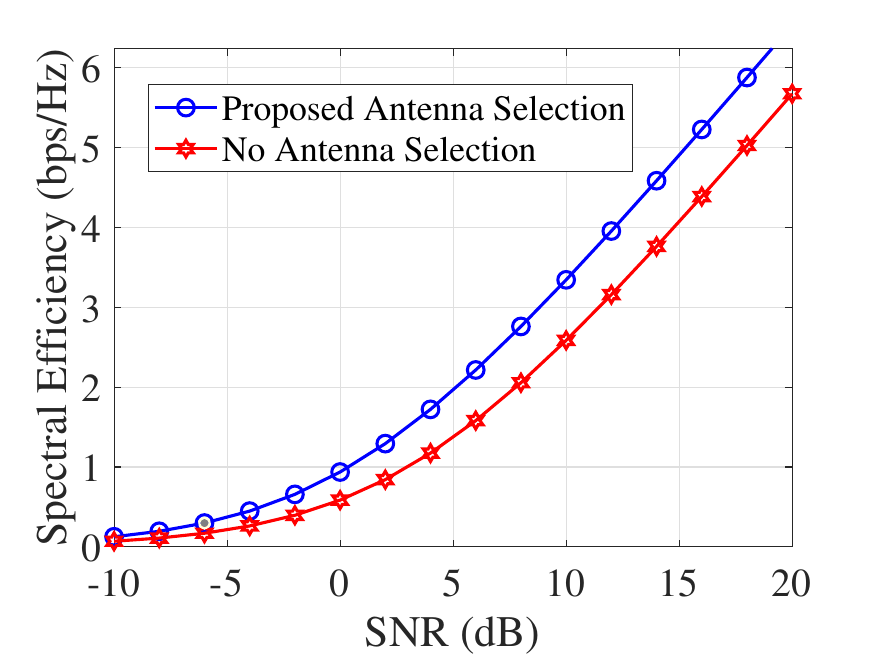}
}
\subfloat[]{
  \includegraphics[width=41.9mm,height=35mm]{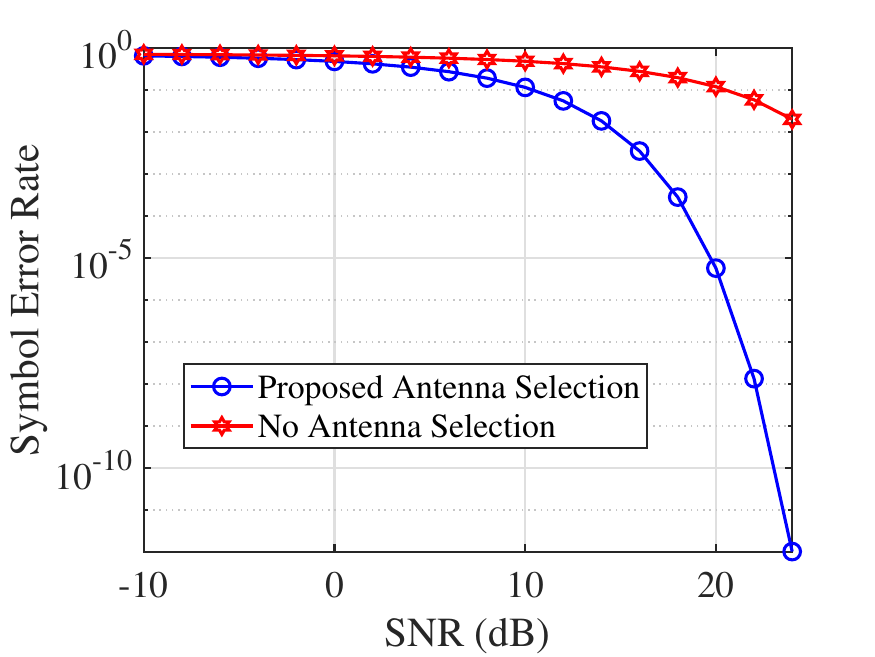}
  }
\caption{(a) \ac{SE} of the proposed \ac{AS} and without \ac{AS}, and (b) \ac{SER} vs \ac{SNR}.}
\label{COMMUNICATION1}
\end{figure}
Figure \ref{sinrComn} shows that the proposed AS enhances \ac{SINR} across varying polarization shifts. At larger shifts (e.g., $60^\circ$ and $80^\circ$), conventional schemes suffer from increased cross-pol interference, while the proposed AS maintains strong co-pol alignment, preventing SINR degradation. SINR is crucial for both sensing and communication, as poor AS directly impacts their performance. To ensure reliable JRC operation, the proposed AS strategically selects antennas based on SINR, optimizing signal reception and target detection accuracy. Fig. \ref{COMMUNICATION1}(a) shows SE performance, where proposed AS consistently outperforms conventional methods. By selecting antennas with optimal co-pol alignment, AS improves throughput and spatial diversity while mitigating interference. Notably, the performance gap widens at high \ac{SNR}, emphasizing AS effectiveness in interference-limited scenarios. Fig. \ref{COMMUNICATION1}(b) shows SER performance, where proposed AS achieves significantly lower error rates. Conventional schemes exhibit an error floor due to suboptimal antennas introducing interference. AS mitigates this issue, ensuring only high-SNR antennas contribute to demodulation, which exponentially reduces SER at high SNR.
\begin{figure*}[ht]
\centering      
\subfloat[]{
  \includegraphics[width=53mm]{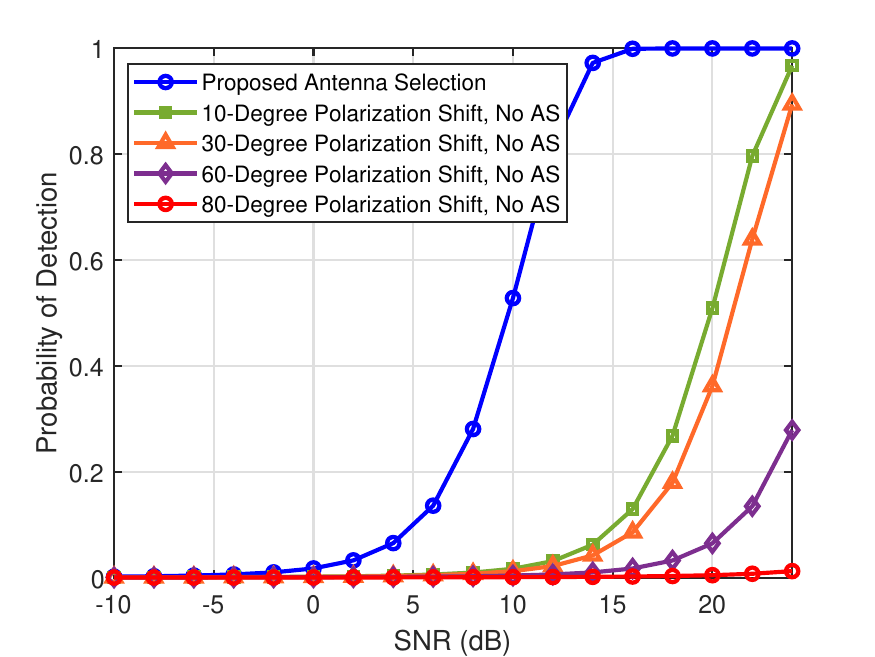}
}
\subfloat[]{
  \includegraphics[width=53mm]{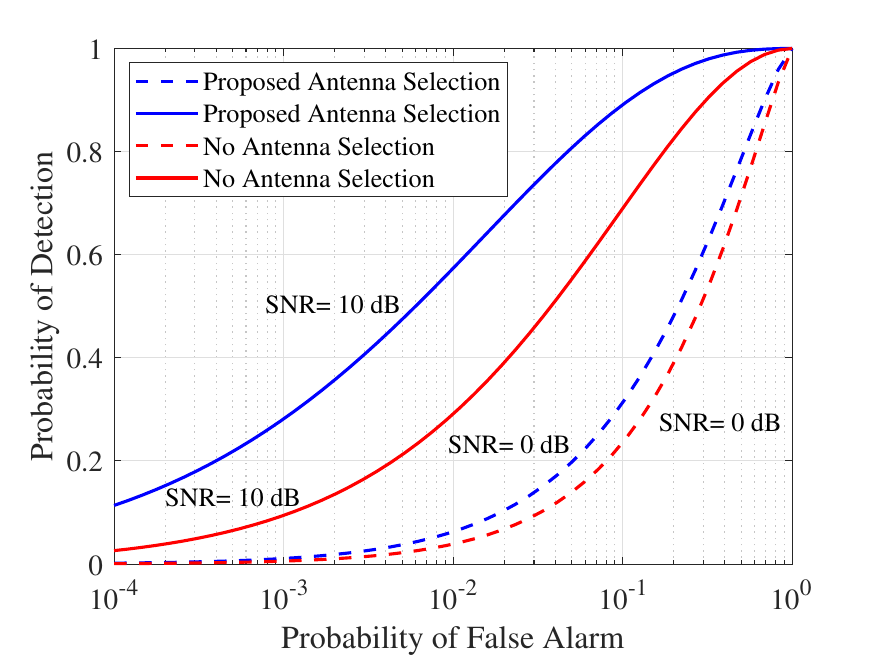}
}
\subfloat[]{
  \includegraphics[width=53mm]{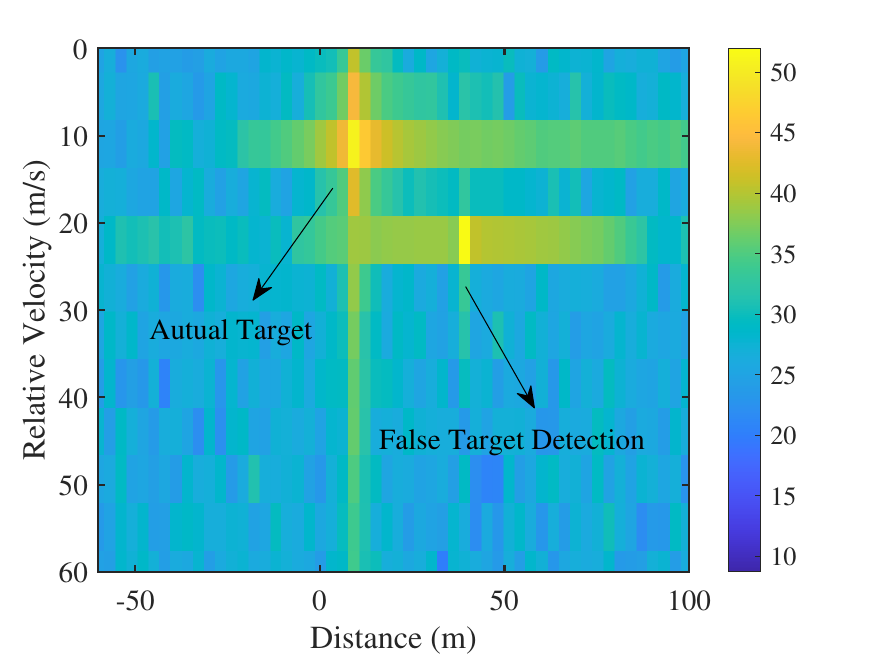}
}
\caption{(a) $P_d$ vs SNR, (b) \ac{ROC} curves for $P_d$ against $P_{fa}$, and (c) velocity-distance plot for the estimated target with and without proposed \ac{AS} scheme.}
\label{radarpart1}
\end{figure*}
\vspace{-5mm}
\subsection{Radar Performance}
Figure \ref{radarpart1}(a) shows that the proposed AS improves $P_d$ across all SNR levels. Without AS, detection performance suffers at large polarization shifts due to reduced effective received power. The proposed AS dynamically selects antennas with strong co-pol alignment, enhancing \ac{$P_d$} while minimizing \ac{$P_{fa}$}. Fig. \ref{radarpart1}(b) presents the ROC curves, where AS achieves superior target detection accuracy, particularly at low $P_{fa}$. Conventional methods exhibit a rightward ROC shift, indicating reduced robustness against multipath-induced interference. The proposed AS ensures reliable detection, a critical requirement for XL-MIMO radar applications. Fig. \ref{radarpart1}(c) further validates the radar sensing improvements via velocity-distance plots. Conventional methods generate false targets due to polarization leakage and multipath interference. The proposed AS mitigates these effects, improving spatial resolution and target separability. This highlights its effectiveness in enhancing radar-based sensing for JRC networks.

\section{Conclusion}
This paper presents a polarization-aware AS scheme for XL-MIMO, optimizing communication and sensing by selecting antennas based on co-pol and cross-pol power levels. The method mitigates polarization leakage and cross-pol interference, improving SE, SER, and radar detection. Unlike heuristic approaches with high complexity, the proposed VR-based selection ensures scalability for large-scale XL-MIMO. Simulations confirm superior performance over conventional methods, making it a strong candidate for next-generation JRC networks. Future work could explore real-time polarization tracking and machine learning-driven adaptive AS to enhance robustness in dynamic environments. Furthermore, the proposed AS can be extended by integrating it with adaptive beamforming to further enhance SE and $P_d$.



\begin{thebibliography}{10}
\providecommand{\url}[1]{#1}
\csname url@samestyle\endcsname
\providecommand{\newblock}{\relax}
\providecommand{\bibinfo}[2]{#2}
\providecommand{\BIBentrySTDinterwordspacing}{\spaceskip=0pt\relax}
\providecommand{\BIBentryALTinterwordstretchfactor}{4}
\providecommand{\BIBentryALTinterwordspacing}{\spaceskip=\fontdimen2\font plus
\BIBentryALTinterwordstretchfactor\fontdimen3\font minus \fontdimen4\font\relax}
\providecommand{\BIBforeignlanguage}[2]{{%
\expandafter\ifx\csname l@#1\endcsname\relax
\typeout{** WARNING: IEEEtran.bst: No hyphenation pattern has been}%
\typeout{** loaded for the language `#1'. Using the pattern for}%
\typeout{** the default language instead.}%
\else
\language=\csname l@#1\endcsname
\fi
#2}}
\providecommand{\BIBdecl}{\relax}
\BIBdecl

\bibitem{wang2023extremely}
Z.~Wang \emph{et~al.}, ``Extremely large-scale {MIMO}: Fundamentals, challenges, solutions, and future directions,'' \emph{IEEE Wireless Commun.}, 2023.

\bibitem{cui2022near}
M.~Cui \emph{et~al.}, ``Near-field {MIMO} communications for {6G}: Fundamentals, challenges, potentials, and future directions,'' \emph{IEEE Commun. Mag.}, vol.~61, no.~1, pp. 40--46, 2022.

\bibitem{de2020non}
E.~De~Carvalho \emph{et~al.}, ``Non-stationarities in extra-large-scale massive {MIMO},'' \emph{IEEE Wireless Commun.}, vol.~27, no.~4, pp. 74--80, 2020.

\bibitem{svantesson2001mutual}
T.~Svantesson and A.~Ranheim, ``Mutual coupling effects on the capacity of multielement antenna systems,'' in \emph{IEEE International Conference on Acoustics, Speech, and Signal Processing. Proceedings}, vol.~4, Salt Lake City, UT, May 2001, pp. 2485--2488.

\bibitem{yuan2021electromagnetic}
S.~S. Yuan \emph{et~al.}, ``Electromagnetic effective degree of freedom of an {MIMO} system in free space,'' \emph{IEEE Antennas Wireless Propag. Lett.}, vol.~21, no.~3, pp. 446--450, 2021.

\bibitem{10458884}
A.~Naeem \emph{et~al.}, ``Polarization-based multiplexing: Enabling spectrum efficient joint radar and communication,'' \emph{IEEE Wireless Commun. Lett.}, vol.~13, no.~5, pp. 1414--1418, 2024.

\bibitem{zheng2019radar}
L.~Zheng \emph{et~al.}, ``Radar and communication coexistence: An overview: A review of recent methods,'' \emph{IEEE Signal Process. Mag.}, vol.~36, no.~5, pp. 85--99, 2019.

\bibitem{zheng2017adaptive}
L.~Zheng, M.~Lops, and X.~Wang, ``Adaptive interference removal for uncoordinated radar/communication coexistence,'' \emph{IEEE J. Sel. Topics Signal Process.}, vol.~12, no.~1, pp. 45--60, 2017.

\bibitem{li2019interference}
Y.~Li, L.~Zheng, M.~Lops, and X.~Wang, ``Interference removal for radar/communication co-existence: The random scattering case,'' \emph{IEEE Trans. Wireless Commun.}, vol.~18, no.~10, pp. 4831--4845, 2019.

\bibitem{deng2013interference}
H.~Deng and B.~Himed, ``Interference mitigation processing for spectrum-sharing between radar and wireless communications systems,'' \emph{IEEE Trans. Aerosp. Electron. Syst.}, vol.~49, no.~3, pp. 1911--1919, 2013.

\bibitem{ciuonzo2016intrapulse}
D.~Ciuonzo \emph{et~al.}, ``Intrapulse radar-embedded communications via multiobjective optimization,'' \emph{IEEE Trans. Aerosp. Electron. Syst.}, vol.~51, no.~4, pp. 2960--2974, 2016.

\bibitem{liu2019interfering}
F.~Liu \emph{et~al.}, ``Interfering channel estimation for radar and communication coexistence,'' in \emph{IEEE 20th International Workshop on Signal Processing Advances in Wireless Communications}, Cannes, France, July 2019, pp. 1--5.

\bibitem{mahal2017spectral}
J.~A. Mahal \emph{et~al.}, ``Spectral coexistence of {MIMO} radar and {MIMO} cellular system,'' \emph{IEEE Trans. Aerosp. Electron. Syst.}, vol.~53, no.~2, pp. 655--668, 2017.

\bibitem{li2017joint}
B.~Li and A.~P. Petropulu, ``Joint transmit designs for coexistence of {MIMO} wireless communications and sparse sensing radars in clutter,'' \emph{IEEE Trans. Aerosp. Electron. Syst.}, vol.~53, no.~6, pp. 2846--2864, 2017.

\bibitem{liu2018mimo}
F.~Liu \emph{et~al.}, ``{MIMO} radar and cellular coexistence: A power-efficient approach enabled by interference exploitation,'' \emph{IEEE Trans. Signal Process.}, vol.~66, no.~14, pp. 3681--3695, 2018.

\bibitem{rao2020probability}
R.~M. Rao, V.~Marojevic, and J.~H. Reed, ``Probability of pilot interference in pulsed radar-cellular coexistence: Fundamental insights on demodulation and limited {CSI} feedback,'' \emph{IEEE Commun. Lett.}, vol.~24, no.~8, pp. 1678--1682, 2020.

\bibitem{10520715}
D.~Galappaththige \emph{et~al.}, ``Near-field {ISAC}: Beamforming for multi-target detection,'' \emph{IEEE Wireless Commun. Lett.}, vol.~13, no.~7, pp. 1938--1942, 2024.

\bibitem{10579914}
H.~Li \emph{et~al.}, ``Near-field integrated sensing, positioning, and communication: A downlink and uplink framework,'' \emph{IEEE J. Sel. Areas Commun.}, vol.~42, no.~9, pp. 2196--2212, 2024.

\bibitem{10772413}
Y.~Lin \emph{et~al.}, ``Near-field integrated sensing and communication beamforming considering complexity,'' \emph{IEEE Trans. Veh. Technol.}, pp. 1--14, 2024.

\bibitem{9737357}
F.~Liu \emph{et~al.}, ``Integrated sensing and communications: Toward dual-functional wireless networks for {6G} and beyond,'' \emph{IEEE J. Sel. Areas Commun.}, vol.~40, no.~6, pp. 1728--1767, 2022.

\bibitem{10437147}
R.~Liu, M.~Li, and Q.~Liu, ``Joint transmit/receive antenna selection and beamforming design for {ISAC} systems,'' in \emph{IEEE Global Communications Conference}, Kuala Lumpur, Malaysia, Dec. 2023, pp. 3118--3123.

\bibitem{7794760}
J.~Yuan \emph{et~al.}, ``A cross-polarization discrimination compensation algorithm for polarization modulation,'' in \emph{IEEE 27th Annual International Symposium on Personal, Indoor, and Mobile Radio Communications}, Valencia, Spain, Dec. 2016, pp. 1--6.

\bibitem{castellanos2023linear}
M.~R. Castellanos and R.~W. Heath, ``Linear polarization optimization for wideband {MIMO} systems with reconfigurable arrays,'' \emph{IEEE Trans. Wireless Commun.}, vol.~23, no.~3, pp. 2282--2295, 2023.

\bibitem{guo2016advances}
C.~Guo \emph{et~al.}, ``Advances on exploiting polarization in wireless communications: Channels, technologies, and applications,'' \emph{IEEE Commun. Surveys Tuts.}, vol.~19, no.~1, pp. 125--166, 2016.

\bibitem{zegrar2022common}
S.~E. Zegrar and H.~Arslan, ``{Common CP-OFDM Transceiver Design for Low-Complexity Frequency Domain Equalization},'' \emph{IEEE Wireless Commun. Lett.}, 2022.

\bibitem{marinello2020antenna}
J.~C. Marinello \emph{et~al.}, ``Antenna selection for improving energy efficiency in {XL}-{MIMO} systems,'' \emph{IEEE Trans. Veh. Technol.}, vol.~69, no.~11, pp. 13\,305--13\,318, 2020.

\bibitem{de2021quasi}
J.~H.~I. de~Souza \emph{et~al.}, ``Quasi-distributed antenna selection for spectral efficiency maximization in subarray switching {XL}-{MIMO} systems,'' \emph{IEEE Trans. Veh. Technol.}, vol.~70, no.~7, pp. 6713--6725, 2021.

\bibitem{6779686}
S.-C. Kwon and G.~L. Stüber, ``Polarization division multiple access on {NLoS} wide-band wireless fading channels,'' \emph{IEEE Trans. Wireless Commun.}, vol.~13, no.~7, pp. 3726--3737, 2014.

\bibitem{9058698}
A.~Behnad, ``A novel extension to craig’s {Q}-function formula and its application in dual-branch {EGC} performance analysis,'' \emph{IEEE Trans. Commun.}, vol.~68, no.~7, pp. 4117--4125, 2020.

\bibitem{10012421}
Z.~Wei \emph{et~al.}, ``Integrated sensing and communication signals toward {5G}-{A} and {6G}: A survey,'' \emph{IEEE Internet Things J.}, vol.~10, no.~13, pp. 11\,068--11\,092, 2023.

\bibitem{delamou2024interference}
M.~Delamou and E.~M. Amhoud, ``Interference reduction design for improved multitarget detection in {ISAC} systems,'' in \emph{IEEE 35th International Symposium on Personal, Indoor and Mobile Radio Communications}, Valencia, Spain, Sept. 2024, pp. 1--6.

\end{thebibliography}
\end{document}